# A Fuzzy-Fuzzy Vault Scheme

Khaled A. Nagaty

The British University in Egypt

El Sherouk City, Suez Desert Road, Cairo, 11837-P.O Box 43.

khaled.nagaty@ bue.edu.eg

***Abstract*:** In this paper an enhanced fuzzy vault scheme is proposed which we refer to as *fuzzy- fuzzy vault* scheme. The proposed scheme builds on the classical fuzzy vault by adding the concept of uncertainty and imprecision to the classical scheme. To lock a secret key $\kappa$ in the classical fuzzy vault the locking and unlocking elements are crisp or real elements and consequently the locking and unlocking operations are strict imperative. In the *fuzzy- fuzzy vault* scheme, Alice locks the secret key $\kappa$ using a set of fuzzy elements that belong to multi-fuzzy set $\tilde{A}$ obtained from a universe public set of fuzzy elements in a multi-fuzzy set $\tilde{F}_q$ and projecting them on polynomial $p$. The elements in multi-fuzzy sets $\tilde{F}_q$ and $\tilde{A}$ are fuzzy- using $m$ membership functions $MF_i, i = 1,2,...m$. Alice selects a set $k$ of fuzzy elements fuzzy- with a specific membership function $MF_k$ from $\tilde{A}$ to lock the vault. To hide the genuine locking points Alice generates a set of fuzzy chaff points that some of them do not lie on polynomial $p$ while the other fuzzy chaff points may lie on polynomial $p$ but fuzzy-with different membership functions other than the membership $MF_k$ used to lock the vault. To unlock the *fuzzy- fuzzy vault* and retrieve the secret key $\kappa$, Bob should have a set of unlocking fuzzy elements belonging to multi-fuzzy set $\tilde{B}$ which substantially overlap with $\tilde{A}$ is required. Then Bob selects $t'_{TF_{ki}}$ fuzzy elements from $\tilde{B}$ which are close to the $t_{TF_k}$ fuzzy elements from $\tilde{A}$ used by Alice to lock the vault. We show that adding uncertainty and imprecision by introducing fuzzy theory will enhance the security threshold of the fuzzy vault.

94A60 Cryptography, 03E72 Fuzzy set theory, 08A72 Fuzzy algebraic structures, 68P25 Data encryption, 68P30 Coding and information theory

## I. INTRODUCTION

The classical fuzzy vault scheme was first introduced by Juels and Sudan in [1]. It is essentially constructed to protect user's secrets whether they are keys or private information by using some features from his/her unique biometrics such as fingerprints. A classical fuzzy vault is composed of a lock algorithm where a set of points are selected to lock the vault and the unlocking algorithm uses a subset that is close enough to the locking set to retrieve the secret information. The classical fuzzy vault includes two parameters the finite field $F_q$ with $q$ a power of prime and the Reed-Solomon error correction algorithm [2]. If we assume that the locking and unlocking of the fuzzy vault has to be made in fuzzy environments, quite a number of fuzzy operations exist. First of all, the user might really not actually want to reduce or minimize the security level of the vault. Rather he might want to reach some higher security levels which might not be achievable using crisp sets for locking and unlocking the classical fuzzy vault. In other words, the user wants to improve the present security levels considerably. Second, the elements and operations might be vague and not behaving as in the strictly mathematical sense but smaller violations are tolerable. The elements, numbers and coefficients can have different membership grades because we assume they are fuzzy in nature or each element or number has fuzzy perception. Consequently the role of the fuzzy elements and fuzzy numbers can be different from that in classical fuzzy vaults where all locking and unlocking elements are of equal weight. In other words, the fuzzy elements of the *fuzzy- fuzzy vault* may be of different importance with different degrees. Fuzzy- fuzzy vaults allow for all types of uncertainty and imprecision and we shall discuss some of them below.

## II. MOTIVATION

The classical fuzzy vaults do not incorporate uncertainty and imprecision in the vault and assume all elements required to lock secret $\kappa$ in the vault or to unlock the vault to retrieve secret $\kappa$ are all of equal importance (weight). Imprecision and uncertainty improve the security threshold of the classical cryptographic techniques. Consequently, locking a secret key under the fuzzy set increases the degree of protection for this key. The concept of chaff points of the fuzzy vault is extended by using membership functions as fuzzy chaff points can lie on polynomial $p$ but have membership grade functions different from the membership grade function used to fuzzify the genuine fuzzy points that lie on polynomial $p$ and used to lock the *fuzzy- fuzzy vault*. This means that fuzzy chaff points will include fuzzy points that do not lie on polynomial $p$ and the fuzzy points that lie on polynomial $p$ but with different membership functions, other than that used in fuzzifying the locking points. Adding this layer of uncertainty to the fuzzy vault and by extending the concept of chaff points will enhance its security level.

This paper is organized as follows: section III presents some preliminaries of fuzzy numbers and fuzzy arithmetic, section IV presents the *fuzzy- fuzzy vault* scheme, section V presents the security analysis of the proposed vault, finally section VI for application of *fuzzy-fuzzy vault* and finally section VII is dedicated for conclusions.

### III. Preliminaries

*A. Fuzzy sets*

If $X$ is a collection of objects denoted generically by $x$, then a fuzzy set $\tilde{A}$ in $X$ is a set of ordered pairs [3]:
$$\tilde{A} = \{(x, \mu_{\tilde{A}}(x) | x \in X)\} \quad (1)$$
Where $\mu_{\tilde{A}}(x)$ is called the membership function which maps $X$ to the membership space $M$. Its range is the subset of nonnegative real numbers whose supremum is finite.
For $\sup \mu_{\tilde{A}}(x) = 1$: normalized fuzzy set.

*B. Fuzzy numbers*

A fuzzy number is a fuzzy set like:

$u: R \rightarrow I = [0, 1]$ which satisfies [4]:
- $u$ is upper semi-continuous,
- $u(x) = 0$ outside some interval $[a, d]$,
- There are real numbers a, b such that a ≤ b ≤ c ≤ d and:
  - $u(x)$ is monotonic increasing on $[a, b]$,
  - $u(x)$ is monotonic decreasing on $[c, d]$,
  - $u(x) = 1, b \leq x \leq c$.

The membership function $u$ can be expressed as:

$$u(x) = \begin{cases} u_L(x), & a \leq x \leq b \\ 1, & b \leq x \leq c \\ u_R(x), & c \leq x \leq d \\ 0, & otherwise \end{cases} \quad (2)$$

Where $u_L: [a, b] \rightarrow [0,1]$ and $u_R: [c, d] \rightarrow [0,1]$ are left and right membership functions of fuzzy number $u$.

1. The trapezoidal fuzzy number $\tilde{a} = (x_0, y_0, \sigma, \beta)$, with two defuzzifiers $x_0, y_0$, and left fuzziness $\sigma > 0$ and right fuzziness $\beta > 0$ is a fuzzy set where the membership function is [5]:

$$\mu_{\tilde{a}}(x) = \begin{cases} \frac{1}{\sigma}(x - x_0 + \sigma) & x_0 - \sigma \leq x \leq x_0 \\ 1, & x \in [x_0, y_0] \\ \frac{1}{\beta}(y_0 - x + \beta) & y_0 \leq x \leq y_0 + \beta \\ 0, & otherwise \end{cases} \quad (3)$$

2. The triangular fuzzy number $\tilde{a} = (a_l, a_s, a_r) \in TF(\tilde{A})$ where $TF(\tilde{A})$ is a triangular membership function of all triangular fuzzy numbers $\tilde{a} \in \tilde{A}$ with membership function $\mu_{\tilde{a}}$ is [4]:

$$\mu_{\tilde{a}}(x) = \begin{cases} 1 - \frac{a_s - x}{a_l}, & (a_s - a_l) \leq x \leq a_s \\ 1 + \frac{a_s - x}{a_r}, & a_s \leq x \leq (a_s + a_r) \\ 0, & otherwise \end{cases} \quad (4)$$

such that $\tilde{a} = (a_l, a_s, a_r) \in TF(\tilde{A})$

3. The sigmoid fuzzy number $\tilde{a} = (a_1, a_2, a_3; \omega)$ in the domain $[-a, a]$ is defined using the membership function [6]:

$$\mu_{\tilde{a}}(x)$$
$$= \begin{cases} \omega\left(\frac{\psi\left(\{x - \frac{a_1 + a_2}{2}\} \times \frac{2a}{a_2 - a_1}\right) - \psi(-a)}{\psi(a) - \psi(-a)}\right), & a_1 \leq x \leq a_2 \\ \omega\left(\frac{\psi(a) - \psi\left(\{x - \frac{a_2 + a_3}{2}\} \times \frac{2a}{a_3 - a_2}\right)}{\psi(a) - \psi(-a)}\right), & a_2 \leq x \leq a_3 \\ 0, & otherwise \end{cases} \quad (5)$$

Where:

$$\psi(a) = \frac{1}{1 + e^{-a}} \quad (6)$$

is a sigmoid function.

4. The Gaussian fuzzy number $\tilde{a}$ has the following membership function [7]:

$$\mu_{\tilde{a}}(x) = \begin{cases} 0, & x \leq \bar{x} - 3\sigma_l \\ \exp\left[-\frac{(x - \bar{x})^2}{2\sigma_l^2}\right], & \bar{x} - 3\sigma_L < x < \bar{x} \\ \exp\left[-\frac{(x - \bar{x})^2}{2\sigma_r^2}\right], & \bar{x} \leq x < \bar{x} + 3\sigma_r \\ 0, & x \geq \bar{x} + 3\sigma_L \end{cases} \quad (7)$$

The mean value is denoted by the parameter $\bar{x}$, and $\sigma_l$ and $\sigma_r$ denoting standard deviations of the Gaussian distribution.

## C. Fuzzy arithmetic

We apply fuzzy arithmetic operations on triangular fuzzy numbers as an example although they can be applied on other types of fuzzy numbers such as sigmoid, Gaussian and trapezoidal fuzzy numbers:

Let $\tilde{a} = (a_l, a_s, a_r), \tilde{b} = (b_l, b_s, b_r) \in TF(R)$ then [8]:

If $x > 0$ then $x\tilde{a} = (xa_l, xa_s, xa_r)$ (8)

If $x < 0$ then $x\tilde{a} = (-xa_l, xa_s, -xa_r)$ (9)

$\tilde{a} + \tilde{b} = (a_l + b_l, a_s + b_s, a_r + b_r)$ (10)

$\tilde{a} - \tilde{b} = (a_l + b_r, a_s - b_s, a_r + b_l)$ (11)

## D. $n^{th}$ Power of fuzzy number

We also apply finding the $n^{th}$ power on triangular fuzzy numbers although it can be applied on other types of fuzzy numbers such as sigmoid, Gaussian and trapezoidal fuzzy numbers:

Let $\tilde{a} = (a_l, a_s, a_r) > 0$ be a fuzzy number, then $\tilde{a}_\alpha = [(a_s - a_l)\alpha + a_l, a_r - (a_r - a_s)\alpha]$ is the α-cut of the fuzzy number $\tilde{a}$ [6].

To calculate the $n^{th}$ power of the fuzzy number $\tilde{a}$ we first take the $n^{th}$ power of the α-cut of $\tilde{a}$ using interval arithmetic:

$(\tilde{a}_\alpha)^n = ([(a_s - a_l)\alpha + a_l, a_r - (a_r - a_s)\alpha])^n$

$= [((a_s - a_l)\alpha + a_l)^n, (a_r - (a_r - a_s)\alpha)^n]$ (12)

To find the membership function $\mu_{\tilde{a}^n}(\tilde{a})$ we equate to $x$ both the first and second component in eq.(12) which gives:

$x = ((a_s - a_l)\alpha + a_l)^n$ (13)

$x = (a_r - (a_r - a_s)\alpha)^n$ (14)

$\sqrt[n]{x} = (a_s - a_l)\alpha + a_l$ (15)

$\sqrt[n]{x} = a_r - (a_r - a_s)\alpha$ (16)

Expressing $\alpha$ in terms of $x$ as follows:

$\alpha = \frac{\sqrt[n]{x} - a_l}{(a_s - a_l)}, \quad a_l^n \leq x \leq a_s^n$ (17)

$\alpha = \frac{a_r - \sqrt[n]{x}}{(a_r - a_s)}, \quad a_s^n \leq x \leq a_r^n$ (18)

Which gives:

$\mu_{\tilde{a}^n}(x) = \begin{cases} \frac{\sqrt[n]{x} - a_l}{(a_s - a_l)}, & a_l^n \leq x \leq a_s^n \\ \frac{a_r - \sqrt[n]{x}}{(a_r - a_s)}, & a_s^n \leq x \leq a_r^n \end{cases}$ (19)

## IV. Fuzzy- Fuzzy Vault Scheme

In this section, we fuzzify the classical fuzzy vault scheme proposed by Ari Juels and Madhu Sudan in [1] by incorporating imprecision and uncertainty through using the fuzzy theory. We show that the *fuzzy- fuzzy vault* scheme will provide a higher security level than the classical fuzzy vault scheme. The main difference between the *fuzzy- fuzzy vault* scheme $\tilde{V}_{\tilde{A}}$ locked using a fuzzy set $\tilde{A}$ and the classical fuzzy vault scheme $V_A$ locked using crisp set $A$ is the fuzzy environment into which the locking and unlocking operations take place. Assume that Alice wants to lock a secret key $\kappa$ under fuzzy set $\tilde{A}$. Alice calculates the key's signature using the CRC-16 bits and append this signature to key $\kappa$. Alice then encodes the augmented secret key $\kappa$ by dividing it into parts and injects these parts into the coefficients of polynomial $p$. Alice selects a finite field $F_q$ with size $q$ and partitions it into $m_F$ subsets with different or equal sizes such that $F_{q_i}, i = 1,2, \dots m_F$. Alice then generates a multi-fuzzy set $\tilde{F}_q$ by fuzzifying its subsets using a set of $m$ membership functions such that:

$MF(\tilde{F}_q) = \bigcup_{i=1}^{m_F} MF_i(\tilde{F}_{q_i})$ where $\tilde{F}_{q_i} \subset \tilde{F}_q$ (20)

Alice selects from $\tilde{F}_q$ a locking fuzzy set $\tilde{A}$ which contains $m_A$ fuzzy subsets with different or equal sizes which are fuzzy- using $m_A$ membership functions such that $\tilde{A}_i \subset \tilde{A}, \forall i = 1,2, \dots m_A$. Alice then forms a multi-fuzzy set of locking points $\tilde{A}$ such that:

$MF(\tilde{A}) = \bigcup_{i=1}^{m_A} MF_i(\tilde{A}_i)$ where $\tilde{A}_i \subset \tilde{A}$ (21)

Where $MF_i(\tilde{A}_i)$ is the fuzzy subset $\tilde{A}_i$ fuzzy- with membership function $MF_i$.

Alice selects a crisp polynomial $p$ such that:

$p(\tilde{a}_i) = \beta_n \tilde{a}_i^n + \beta_{n-1} \tilde{a}_i^{n-1} + \cdots + \beta_1 \tilde{a}_i + \beta_0$ (22)

Where:

$\beta_j$: is the $j^{th}$ coefficient of polynomial $p$.

Alice selects a subset $\tilde{A}_k \subset \tilde{A}$ which is fuzzy- by membership $MF_k$ where $|\tilde{A}_k| = t_{MF_k}$ fuzzy points and project these fuzzy points on polynomial $p$ to obtain the set:

$\{(\tilde{a}_i, p(\tilde{a}_i)) | i = 1 \dots \dots t_{MF_k}\}$ (23)

where:

$\tilde{a}_i \in \tilde{A}_k$, $p(\tilde{a}_i)$ is the projection of $\tilde{a}_i$ on polynomial $p$.

Alice assigns $\tilde{a}_i$ to $\tilde{x}_i$ and assigns $p(\tilde{a}_i)$ to $\tilde{y}_i$ such that:

$R = \{(\tilde{x}_i, \tilde{y}_i) | i = 1 \dots \dots t_{MF_k}\}$ (24)

where:

$(\tilde{x}_i, \tilde{y}_i) \leftarrow (\tilde{a}_i, p(\tilde{a}_i))$

Finally, Alice creates fuzzy chaff points that do not lie on polynomial $p$ or may lie on polynomial $p$ but with membership functions other than the membership function $MF_k$ i.e. $\tilde{u}_i \notin \tilde{A}_{MF_k}$ and $\tilde{v}_i \neq p(\tilde{u}_i)$.

∴ The set of chaff points is $\{(\tilde{u}_i, \tilde{v}_i) | i = 1 \dots t_{MF_{j \neq k}}\}$ (25)

Finally, Alice adds $(\tilde{u}_i, \tilde{v}_i)$ to $R$ so that:
$R = R \cup \{(\tilde{u}_i, \tilde{v}_i)\}$
$\therefore R = \{(\tilde{x}_i, \tilde{y}_i), (\tilde{u}_i, \tilde{v}_i) \mid \forall \tilde{x}_i \in \tilde{A}_k, \forall \tilde{y}_i \in \tilde{F}_q, \tilde{u}_i \notin \tilde{A}_k, \forall \tilde{v}_i \in \{\tilde{F}_q - p(\tilde{u}_i)\}\}$  (26)

If Bob wants to unlock the *fuzzy- fuzzy vault* to retrieve $\kappa$ he must selects a crisp set $B$ and partitions it into $m_B$ subsets. Bob finds $\tilde{B}$ by using a set of $m_B$ membership functions such that:
$MF(\tilde{B}) = \cup_{i=1}^{m_B} MF_i(\tilde{B}_i)$ where $\tilde{B}_i \subset \tilde{B}$  (27)
Then Bob selects a fuzzy subset $\tilde{B}_k$ which is fuzzy- with the membership function $MF_k$ i.e. $\tilde{B}_k \subset \tilde{B}$ where $|\tilde{B}_k| = t_{MF_k}$ which is close enough to subset $\tilde{A}_k \subset \tilde{A}$ where $|\tilde{A}_k| = t_{MF_k}$.
If $|\tilde{a}_i - \tilde{b}_i| \leq \tilde{\delta}$ then Bob adds the corresponding fuzzy point in $R$ to $Q$  (28)
$\therefore Q = \{(\tilde{x}_i, \tilde{y}_i) \mid i = 1 \ldots \ldots t_{MF_k}\}$  (29)
such that:
$\tilde{x}_i \leftarrow \tilde{b}_i, \tilde{y}_i \in R$
If $\tilde{A}_k$ is close enough to $\tilde{B}_k$ then Bob will reconstruct polynomial $p^*$ and reconstruct the secret $\kappa^*$. If the CRC-16 bits signature extracted from $\kappa^*$ is the same as the CRC-16 bits signature appended to $\kappa$ then Bob will obtain the secret key $\kappa$ otherwise he receives *null*.

Now, assume Eve is an attacker who illegally wants to obtain secret $\kappa$ from the *fuzzy- fuzzy vault*. Eve does not know about the crisp polynomial $p$, the fuzzy locking set $\tilde{A}$, the number and size of $\tilde{A}$ subsets, the $m_A$ membership functions used to fuzzify the subsets of $\tilde{A}$, nor the fuzzy subset $\tilde{A}_k \subset \tilde{A}$ used to lock the secret key $\kappa$ in the polynomial $p$. This means that Eve does not have any information on $MF(\tilde{F}_q)$ of eq.(20), $MF(\tilde{A})$ of eq.(21) nor polynomial $p$ in eq.(22). Therefore, Eve must guess a crisp set $B$ and the $m_A$ membership functions to construct $\tilde{B}$ in eq.(27) to be close enough to $\tilde{A}$, guess the fuzzy subset $\tilde{B}_k$ to be close enough to $\tilde{A}_k$, guess polynomial $p$ to find $Q$ in eq.(29).

Assume that Eve has succeeded to guess polynomial $p$, crisp set $B$ but failed to guess the $m_A$ membership functions that are used to fuzzify set $\tilde{A}$, the $m_F$ membership functions that are used to fuzzify the finite field $\tilde{F}_q$, the fuzzy subset $\tilde{B}_k$ to be close enough to $\tilde{A}_k$. Then Eve with a very high probability will fail to reconstruct the secret key $\kappa$ from the *fuzzy- fuzzy vault* as it will be shown in the security analysis in section 5. Without any information about the fuzzy subsets and their membership functions the attacker will not know about the structure of the fuzzy numbers used to lock the vault or the structure of the polynomial $p$ used to hide the secret $\kappa$. For example, assume Alice partitioned the crisp set $F_q$ into four crisp subsets: $F_{q_1}, F_{q_2}, F_{q_3}, F_{q_4}$ and used Gaussian membership function to fuzzify $\tilde{F}_{q_1}$, sigmoid function to fuzzify $\tilde{F}_{q_2}$, trapezoidal function to fuzzify $\tilde{F}_{q_3}$ and triangular membership function to fuzzify $\tilde{F}_{q_4}$ and formed a multi-fuzzy set $\tilde{F}_q = \cup_{i=1}^{4} \tilde{F}_{q_i}$. Assume that Alice has selected three fuzzy subsets $\tilde{A}_1, \tilde{A}_2$ and $\tilde{A}_3$ from $\tilde{F}_q$ such that $\tilde{A}_1$ is fuzzy- with a triangular membership function, $\tilde{A}_2$ is fuzzy- with a Gaussian membership function and $\tilde{A}_3$ is fuzzy- with sigmoid function and she formed $\tilde{A} = \cup_{i=1}^{3} \tilde{A}_i$. She secretly used the fuzzy subset $\tilde{A}_2$ to lock her secret key $\kappa$ into crisp polynomial $p$. So, Eve does not know about the number of subsets of $\tilde{F}_q$ nor the size of each subset and does not know about the $m$ membership functions used to fuzzify these subsets. Also, Eve does not know about the number of subsets of $\tilde{A}$ and the size of each subset and does not know about the membership functions used to fuzzify these subsets. This means that although Eve may guess a correct crisp set $B$ but does not know about the multi-fuzzy sets $\tilde{F}_q$ and $\tilde{A}$ and does not know about the membership functions that used to fuzzify each of them. Also, Eve does not know about the fuzzy subset $\tilde{A}_k$ used to lock the secret key $\kappa$ in polynomial $p$, this makes the fuzzy subset $\tilde{A}_k$ not close enough with any fuzzy subset $\tilde{B}_k$ guessed by Eve. As a result, Eve with a very high probability will fail to obtain the secret key $\kappa$ which adds a better security level to the *fuzzy- fuzzy vault*.

To describe the *fuzzy- fuzzy vault* scheme we need the following definitions:

$F_q$: a finite field of size $q$.
$\tilde{F}_q$: a fuzzy- finite field of size $q$.
$\kappa$: the secret key to be kept into the vault where $\kappa \in F_q^{\kappa}$.
$p(x)$: polynomial of degree less than $\kappa$.
$p(x) \leftarrow \kappa$: assigns secret key $\kappa$ to polynomial $p$.
$\tilde{A} = \{\tilde{a}_1, \tilde{a}_2, \ldots, \tilde{a}_t\}$ is a locking fuzzy set fuzzified with $m_A$ membership functions $MF_j, j = 1, \ldots, m_A$ where $\tilde{a}_i \in MF(\tilde{A})$.
$\tilde{B} = \{\tilde{b}_1, \tilde{b}_2, \ldots, \tilde{b}_t\}$ is an unlocking fuzzy set fuzzified with multiple membership functions $MF_j, j = 1, \ldots, m_B$ where $\tilde{b}_i \in MF(\tilde{B})$.
$t$: is the total number of genuine locking points.
$t_{MF_k}$: is the number of fuzzy- genuine points using membership function $MF_k$.
$\tilde{A}_k$: a locking fuzzy subset of $\tilde{A}$ fuzzified with membership function $MF_k$.
$\tilde{B}_k$: an unlocking fuzzy subset of $\tilde{B}$ fuzzified with membership function $MF_k$
$r - t_{MF_k}$: is the number of chaff points which include fuzzy- genuine points with membership functions other than $MF_k$ and include chaff points that do not lie on polynomial $p$.

$r$: is the total number of genuine points and chaff points such that $r \leq q$.

The FUZZY_LOCK algorithm has an input: secret key $\kappa$, crisp locking set $A$, total number of genuine locking points $t$, total number of locking points and chaff points $r$, finite field $F_q$ and polynomial degree $n$.

---

**FUZZY_LOCK**
**Input:** $\kappa, A, t, r, F_q, n$
$$p(\tilde{x}) = \beta_n \tilde{x}^n + \beta_{n-1} \tilde{x}^{n-1} + \cdots + \beta_1 \tilde{x} + \beta_0$$
**Output:** $\tilde{R} = \{(\tilde{x}_1, \tilde{y}_1), (\tilde{x}_2, \tilde{y}_2), \ldots \ldots (\tilde{x}_r, \tilde{y}_r)\}$
where: $(\tilde{x}_i, \tilde{y}_i) = [(x_{il}, y_{il}), (x_{is}, y_{is}), (x_{ir}, y_{ir})]$
$x_{il}, y_{il}, x_{is}, y_{is}, x_{ir}, y_{ir} \in F_q$
- $(\tilde{x}_i, \tilde{y}_i) \in MF(\widetilde{F_q}) = \cup_{i=1}^{m_F} MF_i(\widetilde{F_{q_i}})$
- $\tilde{X}, \tilde{R} \leftarrow \emptyset$
- **Generate $16-bit$s CRC data from key $\kappa$**
- **Append the $16-bit$s CRC data to key $\kappa$**
- **Divide $\kappa$ into parts.**
- **Inject the $\kappa$ parts into the coefficients of polynomial $p$ so that: $p \leftarrow \kappa$**
- **Select a fuzzy subset $\tilde{A}$ from $\widetilde{F_q}$ such that:**
$TF(\tilde{A}) = \cup_{i=1}^{m_A} MF_i(\tilde{A}_i)$
- **Select fuzzy subset $\tilde{A}_k$ fuzzy- by membership function $MF_k$ from $\tilde{A}$ to lock the vault.**
- **Let $|\tilde{A}_k| = t_{MF_k}$**
- **for $i = 1$ to $t_{MF_k}$ do**
  {
  $(\tilde{x}_i, \tilde{y}_i) \leftarrow (\tilde{a}_i, p(\tilde{a}_i))$
  $\tilde{R} \leftarrow \tilde{R} \cup (\tilde{x}_i, \tilde{y}_i)$
  }
- **for $i = t_{MF_k} + 1$ to $r$ do**
  {
  $\tilde{u}_i \in \widetilde{F_q} - \tilde{A}_k$
  $\tilde{v}_i \in \widetilde{F_q} - \{p(\tilde{u}_i)\}$
  $\tilde{R} \leftarrow \tilde{R} \cup (\tilde{u}_i, \tilde{v}_i)$
  }
- **Scramble the *fuzzy- fuzzy vault* set $\tilde{R}$**

---

The fuzzy set $\tilde{R}$, secret key $\kappa$ and the parameters $n, r$ together are called "*fuzz-fuzzy vault*", denoted by $\tilde{V}_{\tilde{A}}$. In order for Bob to unlock the *fuzz-fuzzy vault* $\tilde{V}_{\tilde{A}}$ to retrieve the secret key $\kappa$, he must choose a crisp set $B$ that is close enough to crisp set $A$ and calculate $\tilde{B}$ using $m$ multiple membership functions using eq.(22). Bob must select a fuzzy subset $\tilde{B}_k$ which is fuzzy- using the membership function $TF_k$. If a fuzzy point $\tilde{b}$ selected from $\tilde{B}_k$ is close enough to a fuzzy point $\tilde{a}$ from $\tilde{A}_k$ then the correspondent point for $\tilde{a}$ in $\tilde{R}$ is added to $\tilde{Q}$. If a significant number of fuzzy points in $\tilde{B}_k$ are close enough to $\tilde{A}_k$ then Bob uses Lagrange interpolation to reconstruct the polynomial $p^*$ and retrieve the reconstructed key $\kappa^*$. If the CRC-16 bits extracted from $\kappa^*$ is the same as the CRC-16 bits appended to $\kappa$ then Bob will retrieve the key $\kappa$ otherwise he retrieves *null*. The FUZZY_UNLOCK algorithm has an input: *fuzzy- fuzzy vault* $\tilde{V}_{\tilde{A}}$ and the crisp unlocking set $B$. Its output is the secret value $\kappa$ if the fuzzy unlocking points are close enough to the fuzzy locking points or null if the fuzzy unlocking points are not close enough to the fuzzy locking points.

---

**FUZZY_UNLOCK**
**Input:** $\tilde{V}_{\tilde{A}}, B$
**Output:** $\kappa \in F_q{}^\kappa \cup \{null\}$
- **Generate $\tilde{B}$ using a set of $m$ membership functions such that:**
$MF(\tilde{B}) = \cup_{i=1}^{m_B} MF_i(\tilde{B}_i)$ where $\widetilde{B_i} \subset \tilde{B}$
- **Select from the fuzzy set $\tilde{B}$ a fuzzy set $\tilde{B}_k$ fuzzified with membership function $MF_k$ to unlock the vault**
- **Let $|\tilde{B}_k| = t_{MF_k}$**
- **for $i = 1$ to $t_{MF_k}$ do**
  {
  $if (|(\tilde{a}_i - \tilde{b}_i|) \leq \tilde{\delta}$ then
  {
  $\tilde{x}_i \leftarrow \tilde{a}_i$
  $\tilde{y}_i \leftarrow$ **Correspondent fuzzy point for $\tilde{a}_i$ in $\tilde{R}$**
  }
  }
- $\tilde{Q} = \{(\tilde{x}_i, \tilde{y}_i) | \forall \tilde{x}_i \in \tilde{B}_k, \forall \tilde{y}_i \in \widetilde{F_q}\}$
- **Use Lagrange interpolation to reconstruct polynomial $p^*$ with degree $n$ which satisfies the points from set $\tilde{Q}$**
- **Retrieve $\beta_i^* \forall i = 0, 1, 2, 3, \ldots n$ for polynomial $p^*$**
- **if $|\beta_i - \beta_i^*| \leq \tilde{\delta}$ then $\beta_i^* = \beta_i$ $\forall i = 0, 1, 2, \ldots n$**
- $\kappa^* =$ **Concatenate** $(\beta_1^*, \beta_2^*, \ldots \ldots \beta_n^*)$
- **Let $m^*$ be the generated $16-bit$s CRC data for the reconstructed key $\kappa^*$**
- **Let $m$ be the extracted 16-bits appended to $\kappa^*$**
- **If $|m - m^*| \leq \delta$ then output $\kappa$**
  **Otherwise *null***

---

If $(|(\tilde{a}_i - \tilde{b}_i|) \leq \tilde{\delta}$ $\forall i = 1, 2, \ldots t_{MF_k}$ then
Assign $\tilde{a}_i$ to $\tilde{x}_i$ and assign $\tilde{y}_i$ to its correspondent fuzzy point for $\tilde{a}_i$ in $\tilde{R}$.

To calculate $p_n(\tilde{x})$ Lagrange interpolation is performed as follows:
$$L_{n,j}(\tilde{x}) = \prod_{k=1, k \neq j}^{n+1} \frac{\tilde{x} - \tilde{x}_k}{\tilde{x}_j - \tilde{x}_k} \qquad (30)$$
Where:
$\{L_{n,j}\}, \forall j = 1, \ldots \ldots n + 1$ are called the Lagrange polynomials for the interpolation fuzzy points $\tilde{x}_1, \tilde{x}_2, \ldots \tilde{x}_{n+1}$ obtained from fuzzy set $\tilde{B}$.
$$\therefore p_n(\tilde{x}) = \sum_{j=1}^{n+1} \tilde{y}_j L_{n,j}(\tilde{x}) \qquad (31)$$

## V. Security Analysis

The security of the proposed *fuzzy- fuzzy vault* scheme depends on:
- The number and size of fuzzy subsets of multi-fuzzy finite field $\widetilde{F}_q$.
- The $m$ membership functions used to fuzzify $\widetilde{F}_q$.
- The number and size of fuzzy subsets of the locking multi-fuzzy set $\tilde{A}$.
- The number and size of the locking fuzzy subset $\tilde{A}_k$.
- The number of fuzzy chaff points $r - t_{MF_k}$ in the set $R$ where $|R| = r$ and $|\tilde{A}_k| = t_{MF_k}$.
- Type and number of fuzzy chaff points $r - t_{MF_k}$ in the set $R$.

As the number of fuzzy chaff points that do not lie on polynomial $p$ or may lie on polynomial $p$ but with different membership functions increase the more fuzziness the *fuzzy- fuzzy vault* to conceal the correct polynomial $p$. As many fuzzy chaff points are added to $R$ the more polynomials $p$ emerge. In the absence of any information about the multi-fuzzy subsets $\tilde{A}$ and $\widetilde{F}_q$ and their $m$ membership functions $MF(\tilde{A})$ and $MF(\widetilde{F}_q)$ an attacker could not know about the structure of the fuzzy points used in locking the vault and cannot distinguish between the large numbers of polynomials to find the correct polynomial $p$. This improves the security threshold of the *fuzzy- fuzzy vault* scheme. The following lemma proves that with high probability many polynomials of degree less than $k$ agree with the target set $R$ in $t_{TF_k}$ places which means there are many spurious polynomials. This lemma is a modification of Lemma 4 presented in [1].

### A. Lemma 1

From Lemma 4 in [1], for every $\mu$, where $0 < \mu < 1$ with probability at least $1 - \mu$ the target set $R$ containing $r$ fuzzy points generated by the algorithm FUZZY_LOCK on polynomial $p$ and a locking fuzzy set $\tilde{A}$ divided into $m_A$ subsets fuzzy- with $m_A$ membership functions as in eq.(21) satisfy the following condition: there exist at least $\mu N$ polynomials $p'$ of degree less than $k$ such that $R$ includes exactly $t_{MF_j}$ fuzzy points of the form $(\tilde{a}, p'(\tilde{a})) \in R$ and $N$ is calculated as follows:

$$N = q^k \binom{r}{t_{MF_j}} \left(\frac{m_A}{q}\right)^{k-t_{MF_j}} \left(1 - \frac{m_A}{q}\right)^{r-t_{MF_j}} \quad (32)$$

Where:

$$p(\tilde{a}_i | MF_j) = \frac{p(\tilde{a}_i \cap MF_j)}{p(MF_j)} = \frac{\frac{1}{q}}{\frac{1}{m_A}} = \frac{1}{q} \times m_A = \frac{m_A}{q} \quad (33)$$

$m_A$: is the number of membership functions used to fuzzify the locking set $\tilde{A}$.
$t_{MF_j}$: is the number of points fuzzy- with the membership function $MF_j$.

Thus the expected number of polynomials of degree less than $k$ that fit with $t_{MF_j}$ of the $r$ random points is $N$. The algorithm FUZZY_LOCK will use the set of $t_{MF_j}$ fuzzy-points with membership function $MF_j$ and fit to polynomial $p$ to output $R$. Then the number of polynomials in agreement with the $t_{MF_j}$ points is much less than $\mu N$. To prove this, construct a huge $(m + 1)$-partite graph where one partite represents polynomials $p$ with degree less than $k$, one vertex for each polynomial. The other $m$-partite vertices represent vectors correspond to $m_A$ fuzzy subsets with $m_A$ membership functions such that:

$$\widetilde{F}^t_q = \bigcup_{j=1}^{m_A} \tilde{y}_{MF_j} \quad (34)$$

where:

$\tilde{y}_{MF_j} = \left(\tilde{y}_{1_{MF_j}}, \tilde{y}_{2_{MF_j}}, \ldots, \tilde{y}_{t_{MF_j}}\right)$ each partite corresponds to vectors with specific membership grade function $MF_j$.

$$\therefore \widetilde{F}^r_q = \widetilde{F}^t_q \cup \widetilde{F}^{r-t}_q \quad (35)$$

Two vertices $p$ and $\tilde{y}_{TF_j}$ are adjacent if $p$ agrees with $\tilde{y}_{MF_j}$ i.e. if $p(\tilde{x}_i) = \tilde{y}_{i_{TF_j}}$ for exactly $t_{TF_j}$ choices of $t$. It is clear that the number of polynomials in agreement with the $t_{MF_j}$ points is much less than the number of polynomials in agreement with $t - t_{MF_j}$. Since it is proved form Lemma 4 in [1] that the number of polynomials in agreement with $t$ points in classical fuzzy vault scheme to be less than $\mu N$, is at most $\mu$. Therefore, as the number of polynomials in agreement with $t_{MF_j}$ in the *fuzzy- fuzzy vault* scheme is much less than $t - t_{MF_j}$ then the number of polynomials that agree with exactly choice of $t_{MF_j}$ points will be much less than $\mu N$ which is the number of polynomials in agreement with $t$ in the classical fuzzy vault scheme with a probability much less than $\mu$. This increases the security of the *fuzz-fuzzy vault* scheme over the *fuzz-fuzzy vault* scheme. The proposed *fuzzy- fuzzy vault* scheme deals better with non-uniform distribution of the fuzzy- locking points as different fuzzy points are fuzzy- with different membership functions and these fuzzy points may not be equally likely. Assume that all fuzzy locking elements in $\tilde{A}$ are fuzzy- with one membership function $MF_j$ then they are all equally likely to come from the same family of fuzzy sets $\tilde{\varepsilon} \subset 2^{\widetilde{U}} = 2^{\widetilde{F}_q}$.

### B. Lemma 2

According to Lemma 5 presented in [1] f*or* every $\mu > 0$, with probability at least $1 - \mu$ the target $R$ generated by the FUZZY_LOCK algorithm to commit to polynomial $p$ with locking set $\tilde{A}$ there exist at least

$$\mu|\tilde{\varepsilon}| q^k \left( \binom{r}{t_{MF_j}} \bigg/ \binom{q}{t_{MF_j}} \right) (m_A/q)^{k-t_{MF_j}} \left(1 - m_A/q\right)^{r-t_{MF_j}} \quad (36)$$

polynomials $p' \in P$ such that $\tilde{R}$ agrees with $p'$ on some subset of $t$ points in the $\tilde{\varepsilon}$ family.

*C. Lemma 3*

From the construction of the *fuzzy- fuzzy vault* scheme it is clear that its security is based on the security of the classical fuzzy vault scheme and on the multi-fuzzy sets which are fuzzy- with multiple fuzzy membership functions. The probability for an attacker to learn about the number of $\tilde{A}$ fuzzy subsets chosen from the subsets of the fuzzy finite field $\tilde{F}_q$ is $\frac{m_A}{m_F}$, also the probability for an attacker to learn about the fuzzy subset $\tilde{A}_j$ fuzzy- by $MF_j$ chosen from $\tilde{A}$ is $\frac{t_{MF_j}}{t}$ where $|\tilde{A}_j| = t_{MF_j}$ and $|\tilde{A}| = t$. The more the chaff points to create noise in set $R$ to hide the genuine polynomial $p$ the more spurious polynomials which look like the genuine polynomial $p$ are generated to protect polynomial $p$ from an attacker. Based on the probability presented in [2], the modified probability for an attacker to obtain the genuine polynomial $p$ with $n$ degree in the *fuzzy- fuzzy vault* is:

$$\prod_{i=0}^{n-1} \left( \frac{m_A}{m_F} \times \frac{t_{MF_j}}{t} \times \frac{t}{r} \right)^i = \left( \frac{m_A}{m_F} \times \frac{t_{MF_j}}{r} \right)^n \quad (37)$$

It is evident from lemma 2 and lemma 3 that the fuzzy-fuzzy vault which is the fuzzy version of the classical fuzzy vault will significantly reduce the probability for an attacker to restore the genuine polynomial $p$. This is much better compared with the probability introduced by the classical fuzzy vault.

## VI. Application of Fuzzy- Fuzzy Vaults

In the example mentioned in [1] where the movie lover's problem is expected to choose 2 movies out of 10 categories and each category contains 1000 movies. The distribution set will be $(\binom{10^3}{2})^{10}$. Suppose that $r = q = 10^4$, $t = 20, k = 16$ then by lemma 5 of the classical vault in [1] one expects to find $2^{106}$ polynomials of degree at most 15 agreeing with data on 20 points. Then with probability is $1 - 2^{-53}$ there will be $2^{53}$ polynomials exhibiting this behavior. Now, Let the movie lover selects five triangular membership functions with different parameters i.e. $m_A = 5$. Assume the movie lover used the 5 fuzzify membership functions to fuzzify all the genuine and chaff points, then by using lemma 2 in this paper then one expects to find $2^{249}$ polynomials of degree at most 15 with data on 20 points. The probability is $1 - 2^{-125}$ for $2^{125}$ polynomials to exhibit this behavior. Thus, we achieved roughly 125-bit security using the *fuzzy- fuzzy vault*. This is much better than 53-bit security using the classical fuzzy vault. Let us use $r = q = 10^4$, $t = 22, k = 18$ then by using lemma 5 in [1] we expect $2^{139}$ polynomials of degree less then 18 agreeing with 22 points out of $10^4$. Then with the probability $1 - 2^{-70}$ there will be $2^{70}$ polynomials exhibiting this behavior. This is equivalent to roughly 70-bit security. Using the proposed *fuzzy- fuzzy vault* with $m_A = 5$ we expect $2^{276}$ polynomials of degree less than 18 with data on 22 points. Then with the probability $1 - 2^{-138}$ there will be $2^{138}$ polynomials exhibiting this behavior. This is equivalent to roughly 138-bit security which is much better than using the classical fuzzy vault. To construct a strong personal entropy system the user in [1] was asked to answer 29 questions correctly out of 32 i.e. $t = 32$ and $k = 25$ thus achieving 85-bit security. By using the *fuzzy-fuzzy vault* scheme we much better than the strong personal entropy system mentioned in [1] with $k=18$ much less than 25.

## VII. Conclusions

The proposed *fuzzy- fuzzy vault* scheme is considered an addition to the community of fuzzy cryptography where imprecision and uncertainty improves the security threshold of the classical cryptographic techniques. As a result the *fuzzy- fuzzy vault* needs less number of genuine points to lock a secret key in the vault and generates a larger number of chaff points which adds noise to conceal the real polynomial which hides the secret key. This will improve the security threshold of the fuzzy vault. The applications of *fuzzy- fuzzy vault* are the same for classical fuzzy vault scheme but with an improved security threshold. An important area for further research is to improve the *fuzzy- fuzzy vault* scheme by using type-2 fuzzy sets.

## VIII. Fingerprints Fuzzy-Fuzzy Vault

Classical fingerprints fuzzy vaults are used to secure data using minutiae. However, the classical fingerprint fuzzy vault assumes that each minutia has an exact orientation which lies in the following set of orientations $\theta°$ and this is not always true. For example the orientation of minutia $m$ can lie in the interval $[\theta°_1, \theta°_2]$ which are taken from the following set of orientations:

$\theta° = \{0°, 22.5°, 45°, 67.5°, 90°, 112.5°, 135°, 157.5°, 180°, 202.5°, 225°, 247.5°, 270°, 292.5°, 315°\}$

Fig. 2 shows the bifurcation and ridge ending minutiae with their orientations.

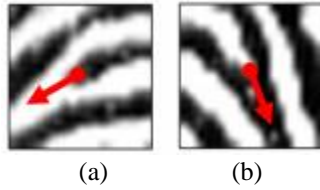

(a)   (b)

Fig. Ridge ending and bifurcation orientations

This means that it a minutia can lie between within an interval of orientations. For example, the ridge ending in (a) can lie in the interval $[202.5°, 225°, 247.5°]$ and the bifurcation in (b) can lie in the interval $[292.5°, 315°, 0°]$. This means that each minutia orientation can be treated as a triangular fuzzy number.